\documentclass[conference]{IEEEtran}
\IEEEoverridecommandlockouts
\usepackage{cite}
\usepackage{amsmath,amssymb,amsfonts,amsthm}
\usepackage[ruled,vlined,linesnumbered]{algorithm2e}
\SetKwIF{If}{ElseIf}{Else}{if}{}{else if}{else}{} 
\SetKwFor{For}{for}{}{}
\SetKwFor{While}{while}{}{}
\SetKw{Continue}{continue}
\SetKw{Break}{break}
\SetKw{Return}{return}
\SetKwProg{Proc}{}{}{}
\SetAlgoLined
\DontPrintSemicolon
\usepackage{graphicx}
\usepackage{textcomp}
\usepackage[monochrome]{xcolor} 
\usepackage[colorlinks=false,allbordercolors={0 0 0},pdfborderstyle={/S/U/W 1}]{hyperref} 
\usepackage{booktabs}
\usepackage{listings}
\usepackage{enumitem}
\usepackage[disable]{todonotes} 

\lstdefinelanguage{K}{
  keywords={rule, syntax, configuration, requires, ensures, andBool, orBool, notBool, Int, Bool, String},
  keywordstyle=\bfseries,
  comment=[l]{//},
  commentstyle=\itshape,
  string=[s]{"}{"},
  stringstyle=\ttfamily,
  basicstyle=\footnotesize\ttfamily,
  breaklines=true,
  breakatwhitespace=true,
  tabsize=2,
  showstringspaces=false,
  columns=flexible,
  moredelim=[s][\bfseries]{<}{>}, 
  moredelim=[s][\itshape]{[}{]}, 
  morecomment=[s]{/*}{*/}
}

\usepackage{subcaption}
\usepackage{multirow}
\usepackage{tabularx}
\usepackage[utf8]{inputenc}
\usepackage[T1]{fontenc}

\newtheorem{theorem}{Theorem}

\newtheorem{definition}[theorem]{Definition}

\newcommand{\K}{{\mathbb{K}}}

\def\BibTeX{{\rm B\kern-.05em{\sc i\kern-.025em b}\kern-.08em
    T\kern-.1667em\lower.7ex\hbox{E}\kern-.125emX}}

\begin{document}

\makeatletter
\newcommand{\globalcolor}[1]{\color{#1}\global\let\default@color\current@color}
\makeatother
\globalcolor{black} 

\title{Compiling by Proving: Language-Agnostic Automatic Optimization from Formal Semantics}

\author{
\IEEEauthorblockN{Jianhong Zhao\IEEEauthorrefmark{1}\thanks{Also with Runtime Verification Inc. (intern).}, Everett Hildenbrandt\IEEEauthorrefmark{2}, Juan Conejero\IEEEauthorrefmark{2}, Yongwang Zhao\IEEEauthorrefmark{1}}
\IEEEauthorblockA{\IEEEauthorrefmark{1}Zhejiang University\\
Email: zhaojianhong96@gmail.com; zhaoyw@zju.edu.cn}
\IEEEauthorblockA{\IEEEauthorrefmark{2}Runtime Verification Inc.\\
Email: everett.hildenbrandt@runtimeverification.com; juan.conejero@runtimeverification.com}
}


\maketitle
\thispagestyle{plain}

\begin{abstract}
Verification proofs encode complete program behavior—every path, every constraint, every transformation—yet we discard them after checking correctness. We present \emph{compiling by proving}, a paradigm that transforms these proofs into optimized execution rules. By constructing All-Path Reachability Proofs through symbolic execution and compiling their graph structure, we consolidate many semantic rewrites into single rules while preserving correctness by construction. We implement this as a language-agnostic extension to the $\K$ framework. Evaluation demonstrates performance improvements across different compilation scopes: opcode-level optimizations show consistent speedups, while whole-program compilation achieves orders of magnitude greater performance gains.
\end{abstract}


\section{Introduction}

Program analysis has evolved along two parallel tracks: compilers analyze programs to optimize performance, while verifiers analyze programs to prove correctness. These communities have developed sophisticated but separate infrastructures. Compilers build control flow graphs, perform dataflow analysis, and apply pattern-based transformations. Verifiers construct reachability proofs, explore symbolic states, and generate verification conditions. Both invest substantial computation analyzing the same programs, yet their artifacts remain incompatible.

However, this separation is artificial. The reachability proofs constructed during verification encode the most precise program analysis possible---they capture all execution paths, state transformations, and semantic constraints. These proofs contain exactly the information compilers seek through heuristic analysis. By compiling these proofs directly, we demonstrate a path toward unifying verification and optimization: constructing proofs simultaneously generates efficient code.

We realize this vision through \textbf{compiling by proving}, implemented in the $\K$ framework~\cite{SerbanutaRosu2010K,ChenRosu2019}—a language-agnostic system for defining formal semantics and generating language tools. $\K$ uses rewrite rules to define formal semantics, enabling concrete and symbolic execution through it. However, executing programs through many rewrites incurs substantial overhead—each step requires pattern matching and rule application. We address this by transforming verification proofs into optimized execution rules, eliminating rewriting overhead while maintaining correctness by construction. This paper presents our method and demonstrates its effectiveness:
\begin{itemize}[leftmargin=*,itemsep=0pt,topsep=2pt]
\item Method: We transform verification artifacts into optimized rules through: (1) program specifications that define compilation boundaries, (2) All-Path Reachability Proofs (APRPs) that capture execution as directed graphs, (3) proof construction via symbolic execution, and (4) graph transformations that compile proofs into optimized rules
\item Implementation: We implement the approach as a language-agnostic $\K$ framework extension, enabling users to construct specifications for diverse compilation units (opcodes, functions) and generate optimized rules through a Python API.
\item Evaluation: We evaluate on EVM semantics by compiling individual opcodes and MIR semantics by compiling an entire program, demonstrating that performance improvements scale with compilation scope from fine-grained operations to complete program execution.
\end{itemize}

Section~\ref{sec:background} provides background on term rewriting systems and motivates the approach. Section~\ref{sec:approach} formalizes the problem and presents our algorithmic solution with theoretical properties. Section~\ref{sec:evaluation} presents comprehensive experimental evaluation with implementation discussion. Section~\ref{sec:related-work} discusses related work and Section~\ref{sec:conclusion} concludes.

\section{Background and Motivation}
\label{sec:background}

\subsection{Term Rewriting Systems and the $\K$ Framework}

Term Rewriting Systems (TRS)~\cite{BaaderNipkow1998} provide the theoretical foundation for executable formal semantics. In TRS, computation unfolds as sequences of rewrite steps that systematically transform terms according to specified rules. The $\K$ framework~\cite{SerbanutaRosu2010K} extends traditional TRS with configuration-based rewriting, enabling modular specification of complex language semantics. These specifications serve as trusted formal foundations for generating interpreters, symbolic execution engines, and program analysis tools.

A $\K$ specification consists of: (1) \textbf{Configuration}: nested cell structures representing program state, (2) \textbf{Syntax}: language constructs and data types, (3) \textbf{Rules}: semantic rules with side conditions defining operational behavior. From a single $\K$ specification, the framework automatically generates interpreters, symbolic execution engines, and program analysis tools. This ``one definition, many tools'' philosophy, underpinned by matching logic~\cite{ChenLucanuRosu2021b}, eliminates the gap between executable interpreters and formal models that plagues traditional approaches~\cite{Zakowski2021Modular}. However, executing programs through many modular rules incurs substantial overhead—motivating our compilation approach.

\subsection{Motivating Example: Compiling EVM Semantics}

The Ethereum Virtual Machine (EVM) powers smart contracts that secure billions of dollars across decentralized finance protocols. As a stack-based machine with 149 opcodes, the EVM requires rigorous formal semantics—bugs have caused billions in losses, making verification critical. The EVM semantics~\cite{Hildenbrandt2018KEVM} provides this foundation through modular $\K$ rules that precisely capture each opcode's behavior.

\begin{figure}[t]
\begin{lstlisting}[basicstyle=\footnotesize\ttfamily, language=K, frame=lines, framesep=3pt]
// Control flow and validation
rule <k> #next[OP] => #addr[OP] ~> #exec[OP] 
                         ~> #pc[OP] </k>
      <wordStack> WS </wordStack>
  requires notBool (#stackUnderflow(WS, OP)
                 orBool #stackOverflow(WS, OP))

// ADD-specific execution
rule <k> #exec[ADD] => #gas[ADD, W0, W1] 
                           ~> ADD W0 W1 </k> 
      <wordStack> W0 : W1 : WS => WS </wordStack>
rule <k> ADD W0 W1 => W0 +Word W1 ~> #push </k>

// Stack and PC management  
rule <k> W:Int ~> #push => .K </k> 
      <wordStack> WS => W : WS </wordStack>
rule <k> #pc[OP] => .K </k> 
      <pc> PC => PC +Int #widthOp(OP) </pc>
\end{lstlisting}
\caption{Modular $\K$ rules for EVM's ADD opcode (simplified)}
\label{fig:add-modular}
\end{figure}

Consider the \texttt{ADD} opcode: conceptually simple (pop two values, push their sum), yet its correct execution involves stack bounds checking, gas accounting, arithmetic computation, and program counter updates. Figure~\ref{fig:add-modular} shows how modular specifications separate these concerns into distinct rules—enabling maintainability and rapid updates as Ethereum evolves. However, this modularity incurs substantial performance costs: executing this single ADD instruction requires five rule applications. Since smart contracts typically contain thousands of opcodes, each requiring multiple rule applications, symbolic execution becomes prohibitively slow.

To eliminate this overhead while preserving semantic correctness, our \emph{compiling by proving} paradigm automatically consolidates these modular rules. For the ADD example in its typical execution path, our approach compiles the five separate rules into a single optimized rule:

\begin{lstlisting}[basicstyle=\footnotesize\ttfamily, language=K, frame=lines, framesep=3pt]
rule <k> #next[ADD] => .K </k>
     <wordStack> W0 : W1 : WS 
                => (W0 +Word W1) : WS </wordStack>
     <pc> PC => PC +Int 1 </pc>
     <gas> G => G -Int 3 </gas>
  requires #sizeWordStack(W0 : W1 : WS) <=Int 1024
     andBool G >=Int 3
\end{lstlisting}

\section{Approach}
\label{sec:approach}

Our \textit{compiling by proving} paradigm automatically generates optimized rules from formal semantics. This section presents our approach in four parts. First, we formalize program specifications that define compilation boundaries (\S\ref{sec:program-spec}). Next, we introduce our intermediate representation for capturing execution paths (\S\ref{sec:aprp-ir}). Then, we present the algorithm for constructing this representation (\S\ref{sec:aprp-construction}). Finally, we show how to perform proof-based compilation (\S\ref{sec:proof-compilation}).

\subsection{Program Specifications}
\label{sec:program-spec}

Program specifications define compilation boundaries by pairing initial and final states. Each specification delineates a computational unit—an opcode, basic block, or function—through its boundary conditions. To represent these states precisely, we need a formalism that captures both structural configuration and logical constraints.

\begin{definition}[Constrained Term (CTerm)]
\label{def:constrained-term}
A constrained term $\varphi \land C$ combines structural state $\varphi$ (stack, memory, program counter) with logical constraints $C$ (bounds, equalities, invariants). This separation enables precise proof construction: $\varphi$ captures program configuration while $C$ accumulates path conditions during symbolic execution.
\end{definition}

\begin{definition}[Program Specification]
\label{def:program-spec}
A program specification is a pair of constrained terms $(\varphi_{init} \land C_{init}, \varphi_{final} \land C_{final})$ that defines a compilation unit through its initial state and final state pattern. The specification determines the optimization scope: all execution paths from states satisfying $\varphi_{init} \land C_{init}$ that reach states satisfying $\varphi_{final} \land C_{final}$ are compiled into single optimized rules.
\end{definition}

For KEVM opcodes, specifications are straightforward: $\varphi_{init}$ contains the opcode followed by remaining execution, while $\varphi_{final}$ contains only the remaining execution. For the ADD opcode: $\varphi_{init}$ matches $\langle \texttt{ADD} \cdot \mathit{rest} \rangle$ and $\varphi_{final}$ matches $\langle \mathit{rest} \rangle$—capturing all outcomes.

Higher-level specifications follow the same principle. At function level, $\varphi_{init}$ captures function entry with symbolic parameters, while $\varphi_{final}$ identifies function exit with execution returned to the caller.

\subsection{All-Path Reachability Proofs as IR}
\label{sec:aprp-ir}

To compile a program specification into optimized rules, we need an intermediate representation that captures all possible execution paths. All-Path Reachability Proofs (APRPs) serve this purpose, encoding complete execution behavior as a graph structure that makes optimization opportunities explicit.

\begin{definition}[All-Path Reachability Proof]
\label{def:APRP}
An All-Path Reachability Proof $\Pi = (V, E_s, E_c, E_b)$ is a directed graph where vertices $V \subseteq \mathit{CTerm} \times \mathit{Status}$ track execution states during construction and edges capture three types of transitions: steps ($E_s$) for sequential execution, covers ($E_c$) for subsumption, and branches ($E_b$) for conditionals.
\end{definition}

\begin{definition}[Constrained Substitution (CSubst)]
\label{def:csubst}
A constrained substitution $\alpha \equiv (\alpha_s, \alpha_c)$ pairs structural substitution $\alpha_s$ with constraint conjunction $\alpha_c$. Application yields $\alpha(\varphi) \equiv \alpha_s(\varphi) \land \alpha_c$, propagating both state transformations and logical conditions.
\end{definition}

Step and branch edges model program execution, while cover edges enable proof termination. Step edges $(v_1, v_2, n) \in E_s$ capture basic blocks: $n$ deterministic rewrites from $v_1$ to $v_2$ where exactly one rule matches at each state. Branch edges $(v, \{(\alpha_i, v_i)\}) \in E_b$ represent control flow splits: multiple rules match $v$, creating conditional paths to each $v_i$ under guard $\alpha_i$. Cover edges $(v_1, v_2, \alpha) \in E_c$ identify subsumption: when $v_1 = \alpha(v_2)$, concrete state $v_1$ instantiates the general pattern $v_2$.

\subsection{Constructing APRPs through Symbolic Execution}
\label{sec:aprp-construction}

We construct APRPs through deductive proving by applying semantic rules. Starting from the specification's initial state, we build proof trees that establish reachability to the final state through two fundamental operations provided by the K framework's symbolic execution engine:

\noindent\textbf{\texttt{execute($\varphi$, N)}:} Applies up to $N$ semantic rules from state $\varphi$, returning the resulting state $\psi$, branch conditions $[\alpha_i]$ when multiple rules match, and the number of rules applied $M$.

\noindent\textbf{\texttt{implies($\varphi_1$, $\varphi_2$)}:} Proves $\varphi_2$ subsumes $\varphi_1$ through matching logic deduction and SMT solving, returning substitution $\alpha$ when $\alpha(\varphi_2) = \varphi_1$ holds.

The algorithm requires two user-supplied predicates that customize compilation behavior for specific domains:

\noindent\textbf{\texttt{terminal($T$)}:} Returns true if state $T$ represents a valid termination point for the compilation unit (e.g., reaching the final specification state or a stuck state that should end compilation).

\noindent\textbf{\texttt{sameloop($T, T_{prev}$)}:} Returns true if states $T$ and $T_{prev}$ belong to the same iterative structure, enabling targeted abstraction for loop handling.

Using these operations, we construct the complete APRP through the worklist-based algorithm shown in Algorithm~\ref{alg:aprp}. The algorithm offers two modes controlled by the precision flag $P$: fast mode with customized loop detection and precision mode with exhaustive subsumption checking. The iteration bound $I$ ensures guaranteed termination for production deployment, while $N$ limits steps per execution call to control graph granularity.

The algorithm starts by initializing vertices with the initial and final specifications (line 2). For each pending vertex selected from the worklist (line 3), it first checks if the state is terminal using \texttt{terminal($T$)} (line 5). 

For non-terminal states, the algorithm examines previously reached states accessible through $\mathit{reachable}^{\uparrow}$ traversal. In precision mode, it checks for subsumption using \texttt{implies($T, T_{prev}$)} (line 9). For potential loops identified by \texttt{sameloop($T, T_{prev}$)}, it applies the $\mathit{abstract}$ function to create structural generalizations (lines 11-12):
$$\mathit{abstract}(T_1, T_2) := \mathit{cau}(t_1, t_2) \land c_{1 \cap 2}$$
where $\mathit{cau}$ replaces differing subterms with symbolic variables and $c_{1 \cap 2}$ preserves common constraints. This creates weak structural loop invariants ensuring termination while maintaining precision.

If abstraction produces a generalized state, the algorithm adds it as pending and creates a cover edge (line 15). Otherwise, it applies \texttt{execute($T$, $N$)} to advance the symbolic execution (line 18). The execution creates step edges for deterministic sequences (line 21) and branches when multiple rules match, with successors added as pending vertices (lines 22-24).

\begin{algorithm}[t]
\caption{APRP Construction}
\label{alg:aprp}
\SetKwFunction{FMain}{ConstructAPRP}
\SetKwProg{Fn}{procedure}{}{}
\Fn{\FMain{$T_{init}, T_{final}, N, I, P$}}{
    $V \leftarrow \{(T_{init}, \mathit{pending}), (T_{final}, \mathit{final})\}$; $E_s, E_c, E_b \leftarrow \emptyset$; $i \leftarrow 0$\;
    \While{$\exists (T, \mathit{pending}) \in V$ and $i < I$}{
        Select $(T, \mathit{pending}) \in V$; $i \leftarrow i + 1$\;
        \If{$\mathit{terminal}(T)$}{mark $T$ reached; \Continue}
        $T' \leftarrow T$\;
        \For{$(T_{prev}, \_) \in \mathit{reachable}^{\uparrow}(T)$}{
            \If{$P$ and $\mathit{implies}(T, T_{prev}) = \alpha$}{
                add cover $(T, T_{prev}, \alpha)$\;
                mark $T$ reached; \Break;
            }
            \ElseIf{$\mathit{sameloop}(T, T_{prev})$}{
                $T' \leftarrow \mathit{abstract}(T', T_{prev})$\;
            }
        }
        \If{$T' \neq T$ and $\mathit{implies}(T, T') = \alpha$}{
            add $(T', \mathit{pending})$ to $V$, cover $(T, T', \alpha)$\; mark $T$ reached\;
        }
        \If{$T$ is reached \Continue}{}
        $(T_{next}, [\alpha_i], M) \leftarrow \mathit{execute}(T, N)$\;
        \If{$M = 0$}{mark $T$ stuck; \Continue}
        add step $(T, T_{next}, M)$; mark $T$ reached\;
        \If{$|\alpha_i| \geq 1$}{
            add branch $(T_{next}, [(\alpha_i(T_{next}), \alpha_i)])$\;
            add successors $\{\alpha_i(T_{next})\}$ as pending\;
        }
        \Else{add $(T_{next}, \mathit{pending})$ to $V$\;}
    }
    \Return $(V, E_s, E_c, E_b)$\;
}
\end{algorithm}

The algorithm performs deductive proving where each vertex represents a proved reachability claim and each edge records the proof structure. Step edges capture sequences of rule applications, branch edges encode case analysis in the proof, and cover edges establish subsumption lemmas. The algorithm terminates in $O(I \times |V|^2)$ time through either frontier exhaustion or the iteration bound. In practice, KEVM opcodes complete within seconds to minutes depending on complexity (see Table~\ref{tab:rq2_aggregate} for detailed measurements).

\subsection{Proof-Based Compilation}
\label{sec:proof-compilation}

With the APRP constructed for a specification, we transform its graph structure into optimized rules specific to that compilation unit. Each step edge in the APRP represents an optimization opportunity: a sequence of $n$ deterministic rewrites that can be consolidated into a single semantic rule.

Consider a step edge $(v_1, v_2, n)$ capturing the transformation from state $v_1$ to $v_2$ through $n$ rewrites. We compile this into:
$$\text{rule } v_1.\text{state} \Rightarrow v_2.\text{state} \text{ requires } v_1.\text{constraints}$$

Consolidating $n$ rewrites into one rule eliminates rewriting overhead, with performance gains increasing as $n$ grows. This motivates our core strategy: maximize step edge length through graph transformations.

\subsubsection{Graph Transformations for Optimization}

The initial APRP captures the complete execution behavior of the compilation unit. Step edges correspond to basic blocks—deterministic execution sequences—while branches reflect the program's control structure (conditionals, loops). The length of individual step edges is bounded by the parameter $N$ used during construction. We apply semantic-preserving transformations to create longer consecutive steps.

We employ two classes of transformations: step compression for merging deterministic sequences, and branch lifting for restructuring control flow.

\textbf{Step Compression.} This transformation merges consecutive step edges to eliminate intermediate states. Given adjacent edges $v_A \xrightarrow{M} v_B \xrightarrow{N} v_C$, we compose them into $v_A \xrightarrow{M+N} v_C$. The correctness is immediate: applying rules $r_1, \ldots, r_M, r_{M+1}, \ldots, r_{M+N}$ sequentially from $v_A$ produces $v_C$, whether we record the intermediate state $v_B$ or not.

\textbf{Branch Lifting.} These transformations restructure control flow to expose opportunities for step compression.

Step-Branch Lifting moves branches before steps. Given the pattern:
$$v_A \xrightarrow{M} v_B \xrightarrow{[\alpha_1, \ldots, \alpha_n]} [v_{C_1}, \ldots, v_{C_n}]$$
we transform it to:
$$v_A \xrightarrow{[\alpha_1, \ldots, \alpha_n]} [\alpha_1(v_A) \xrightarrow{M} v_{C_1}, \ldots, \alpha_n(v_A) \xrightarrow{M} v_{C_n}]$$

The transformation is valid because the $M$ steps from $v_A$ to $v_B$ are deterministic—no branching occurs. Therefore, applying branch conditions before or after these steps yields identical outcomes.

Branch-Branch Lifting flattens nested conditionals. Given:
$$v_A \xrightarrow{[\alpha_i]} [v_{B_i}] \text{ where } v_{B_k} \xrightarrow{[\alpha_j^k]} [v_{C_j}^k]$$
we produce:
$$v_A \xrightarrow{[\alpha_k \circ \alpha_j^k]} [(\alpha_k \circ \alpha_j^k)(v_A) \rightarrow v_{C_j}^k]$$

The transformation is valid because reaching $v_{C_j}^k$ requires taking branch $k$ from $v_A$ (applying $\alpha_k$) then branch $j$ from $v_{B_k}$ (applying $\alpha_j^k$). The composition $\alpha_k \circ \alpha_j^k$ captures both constraints.

After iterative transformation, the APRP becomes a streamlined graph: from the initial state, branches lead directly to maximally compressed step sequences reaching final states. Each execution path—originally fragmented across many small steps—is now a single optimized rule.

\subsubsection{Correctness by Construction}

Our approach guarantees correctness through three interlocking mechanisms:

First, \textbf{source correctness}: The APRP derives from formal semantics through deductive proving. Each edge is a proved reachability claim established by applying semantic rules.

Second, \textbf{transformation correctness}: Graph transformations preserve semantic equivalence by construction. Step compression eliminates only intermediate states in deterministic sequences. Branch lifting reorders operations without changing execution outcomes.

Third, \textbf{deployment correctness}: Compiled rules are integrated with higher priority than original modular rules~\cite{SerbanutaRosu2010K,ChenLucanuRosu2021b}. The $\K$ framework's priority mechanism ensures that when patterns match, the optimized compiled rules fire first, bypassing multi-step execution while preserving semantic equivalence. Original modular rules remain available with lower priority as fallback, ensuring completeness when compiled rules don't apply.

Through these three correctness mechanisms, our \emph{compiling by proving} paradigm ensures that optimization preserves semantics at every level. This approach demonstrates that verification artifacts need not be mere certificates—they can drive compilation itself. By transforming proof structure into optimized rules, we achieve substantial performance improvements (detailed in \S\ref{sec:evaluation}) while maintaining formal guarantees.

\section{Evaluation}\label{sec:evaluation}

We experimentally evaluate our compiling by proving approach on two complementary case studies: Ethereum Virtual Machine (EVM) opcodes for fine-grained compilation and Rust programs (via MIR) for whole-program compilation.

\subsection{Experimental Setup}

Our evaluation targets two distinct systems. EVM is the stack-based runtime for Ethereum smart contracts, with 149 opcodes handling computation, storage, and contract interactions. For Rust programs, we target MIR (Mid-level Intermediate Representation), Rust's intermediate representation where type checking, borrow checking, and optimizations occur during compilation.

The $\K$ Framework provides formal semantics for EVM, used by Kontrol for production DeFi verification, and formal semantics for MIR enabling formal analysis of Rust programs. These semantics validate our approach across distinct domains (blockchain vs. systems programming) and compilation granularities (individual opcodes vs. complete programs).

For EVM, we constructed program specifications for 131 out of 149 opcodes, evaluating performance with 2756 conformance tests (concrete execution) and 70 symbolic verification tasks. For MIR, we demonstrated whole-program compilation using an iterative sum computation with 1000 loop iterations. Unless stated otherwise, experiments ran on an Apple M2 (8 cores, 16GB RAM) laptop with macOS 15.6, using consistent configurations and standardized timeout limits.

\subsection{EVM Opcode Compilation by Proving}

Table~\ref{tab:step3_category_statistics} summarizes category-level compilation results. We constructed program specifications for 131 out of 149 EVM opcodes (87.9\% coverage). For opcodes with specifications, compilation by proving achieved an average 89.6\% reduction in rewriting steps with 73.3s average compilation time. Categories such as Comparison, Stack, Memory, Storage, and Transactional Storage achieved 100\% specification coverage, while Flow Control (77.8\%) and Context (77.8\%) had lower coverage due to control-flow complexity. Contract-level operations remain future work.

\begin{table}[htbp]
\centering
\caption{Statistics of opcode compilation by category. \#: number of opcodes; Spec. Cov.: proportion of opcodes with successfully constructed program specifications; Time: average compilation by proving time in seconds; $\Delta$ Steps: relative reduction in rewriting steps, defined as $\Delta\text{Steps} = 1 - \frac{\text{steps}_{\text{compiled}}}{\text{steps}_{\text{original}}}$.}
\label{tab:step3_category_statistics}
\begin{tabular}{lrrrr}
\toprule
Category & \# & Spec. Cov. & Time (s) & $\Delta$ Steps \\
\midrule
Arith. \& Bit. & 19 & 84.2\% & 52.0 & 87.5\% \\
Comparison & 6 & 100.0\% & 64.4 & 89.5\% \\
Flow Control & 9 & 77.8\% & 52.1 & 85.4\% \\
Stack & 66 & 100.0\% & 77.4 & 89.2\% \\
Memory & 5 & 100.0\% & 78.3 & 90.1\% \\
Storage & 2 & 100.0\% & 97.9 & 90.5\% \\
Trans. Storage & 2 & 100.0\% & 72.9 & 88.9\% \\
Environment & 9 & 88.9\% & 65.3 & 88.2\% \\
Context & 18 & 77.8\% & 66.6 & 87.2\% \\
System & 6 & 83.3\% & 41.8 & 80.0\% \\
Contract & 7 & 0.0\% & 0.0 & 0.0\% \\
Total & 149 & 87.9\% & 73.3 & 89.6\% \\
\bottomrule
\end{tabular}
\end{table}

The varying specification coverage reflects different challenges in constructing program specifications. Flow Control and Context opcodes require complex termination and initial conditions due to path-sensitive control transfers and environment-dependent side effects. Contract-level operations (e.g., external calls) have non-trivial semantic boundaries due to nested execution contexts and cross-contract state changes, remaining as future work. The complete per-opcode breakdown is provided in Appendix~\ref{sec:opcode-details}.

\begin{figure}[htbp]
	\centering
	\begin{minipage}[t]{0.49\linewidth}
		\centering
		\includegraphics[width=\linewidth]{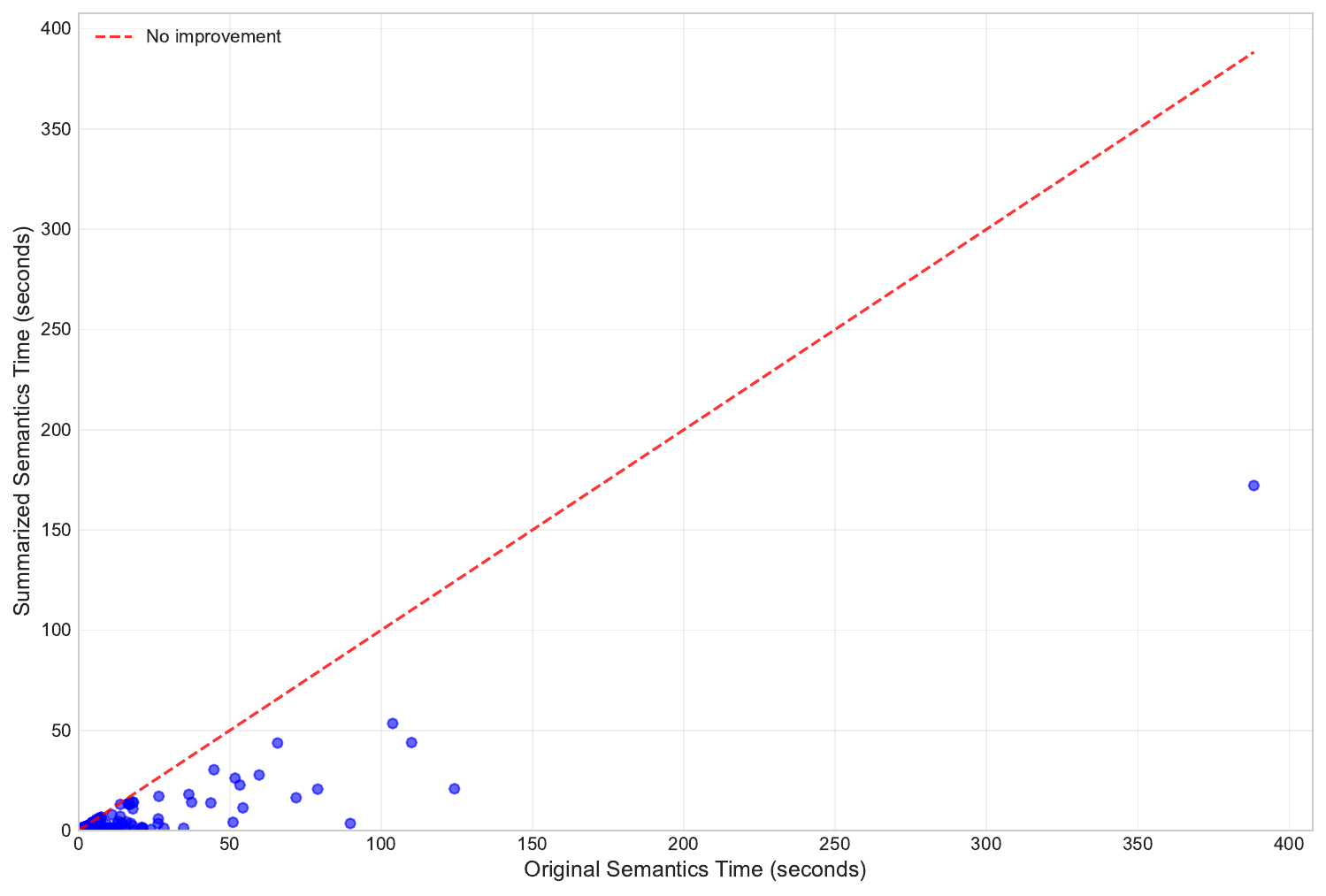}\\[2pt]
		\footnotesize (a) Concrete execution
	\end{minipage}
	\hfill
	\begin{minipage}[t]{0.49\linewidth}
		\centering
		\includegraphics[width=\linewidth]{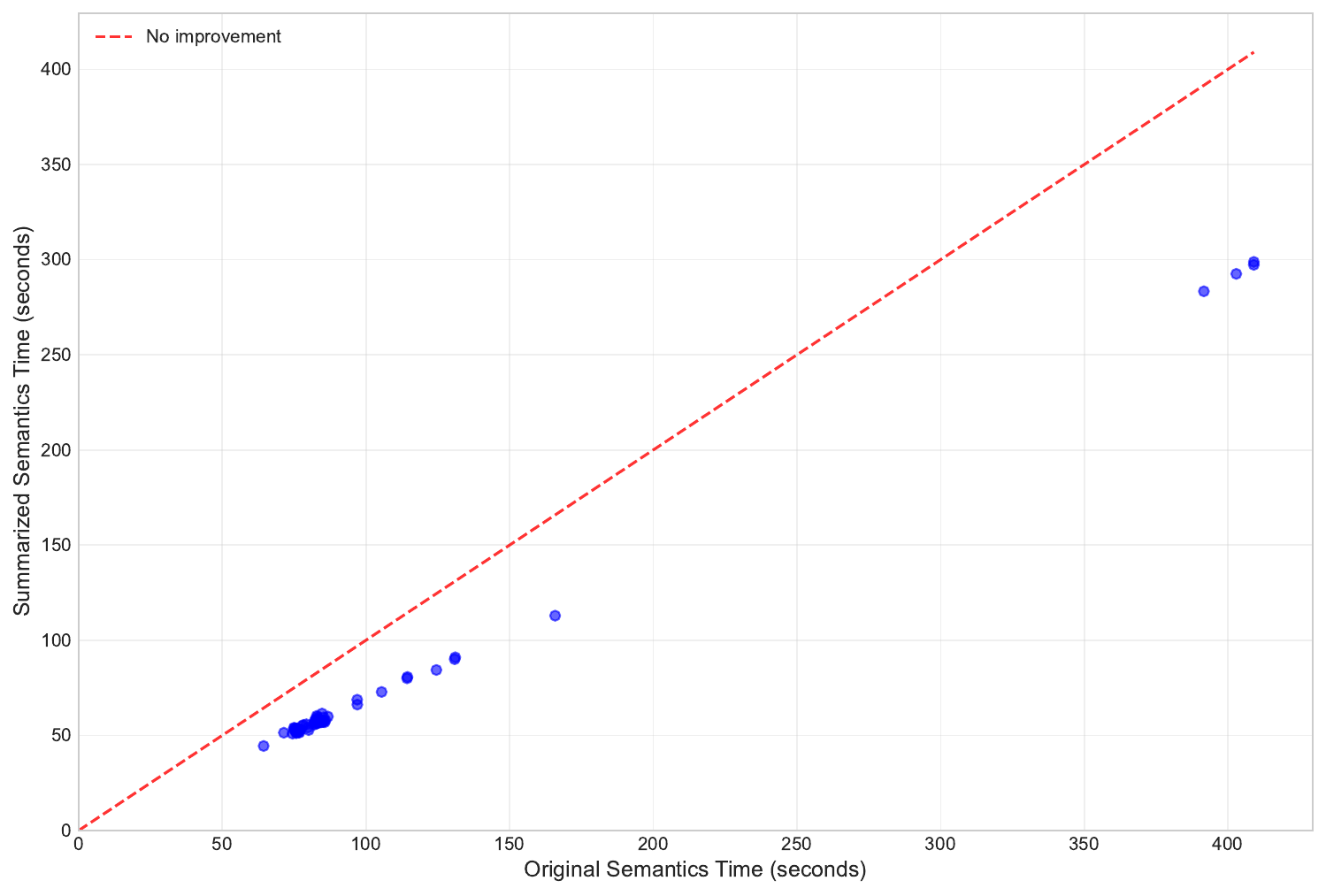}\\[2pt]
		\footnotesize (b) Symbolic execution
	\end{minipage}
	\caption{Original vs. compiled performance.}
	\label{fig:performance-scatter}
\end{figure}

For the 131 opcodes successfully compiled by proving, we integrated their optimized rules into the EVM semantics and evaluated performance against the original semantics. We tested both concrete execution (using the standard EVM conformance suite) and symbolic execution (using compilation-proof tasks) to assess the impact across different execution modes. Figure~\ref{fig:performance-scatter} shows per-test scatter plots comparing original vs. compiled performance, where points below the diagonal indicate speedups. Table~\ref{tab:rq2_aggregate} presents aggregate statistics.

\begin{table}[htbp]
	\centering
	\caption{Aggregate performance improvements (paired, compiled vs. original). Geomean and median speedups are computed on matched test pairs; Wins/Ties/Losses count pairs with speedup $>$ 1.0, $\approx$ 1.0 (\(\pm\)1\%), and $<$ 1.0, respectively.}
	\label{tab:rq2_aggregate}
	\begin{tabular}{lrrrrr}
		\toprule
		Scenario & N    & Geomean & Median & p90   & Wins/Ties/Losses \\
		\midrule
		Concrete & 2756 & 2.189   & 2.000  & 5.286 & 2483/269/4       \\
		Symbolic & 70   & 1.442   & 1.448  & 1.483 & 70/0/0           \\
		\bottomrule
	\end{tabular}
\end{table}

The compiled semantics achieve performance improvements across both execution modes. In concrete execution with 2756 conformance tests, we observe a geometric-mean speedup of 2.19$\times$ (median 2.00$\times$), with the top decile reaching 5.29$\times$. Symbolic execution shows consistent improvements with a 1.44$\times$ speedup (median 1.45$\times$) across 70 proof tasks, with no performance regressions.

These speedups result from eliminating intermediate rewriting steps in the compiled rules. While individual opcode optimizations demonstrate clear benefits, the true potential of compilation by proving emerges when applied to complete programs, as we show next with MIR semantics.

\subsection{Rust Program Compilation by Proving}

While EVM opcode compilation demonstrates operation-level benefits, we now evaluate whole-program compilation using the MIR semantics. Unlike the fine-grained EVM optimizations, whole-program compilation can eliminate the cumulative overhead of thousands of semantic transitions.

We evaluated a representative Rust program that iteratively increments a counter using function calls:
\begin{lstlisting}[basicstyle=\footnotesize\ttfamily, keywords={fn,let,mut,while,assert}, keywordstyle=\bfseries, frame=lines, framesep=3pt, aboveskip=0.5em, belowskip=0.5em]
fn add1(x: i32) -> i32 {
    x + 1
}

fn main() {
    let mut x = 0;
    let mut i = 0;
    while i < 1000 {
        x = add1(x);
        i = i + 1;
    }
    assert!(x == 1000);
}
\end{lstlisting}
This program performs 1000 loop iterations with 1000 function calls and 2000 increment operations, resulting in thousands of MIR-level semantic transitions.

Our compilation by proving consolidates the entire program's semantic rules into a single optimized execution rule. We measure performance through symbolic execution, where each step must explore all semantic rules—making it particularly sensitive to our optimization's benefits. Without compilation, symbolic execution requires 2438.26 seconds. With our compiled rule, it completes in 4.668 seconds—achieving a \textbf{522×} speedup. This speedup, two orders of magnitude greater than our EVM results, demonstrates how compilation benefits scale with program complexity.

\subsection{Discussion}

Beyond performance improvements, the compiled rules enable practical applications such as cross-model validation. Using the $\K$ framework's Lean 4 backend, we compile our optimized EVM rules into Lean 4 and prove their equivalence to Nethermind's formal EVM model. The compilation simplifies cross-framework verification by reducing complex multi-rule semantics to single consolidated rules.

Our compilation framework, implemented in Python, separates language-specific details from generic compilation logic. Users provide program specifications and define terminal conditions for their domain, while the compilation algorithms work uniformly across languages. This design enables the same pipeline to optimize both EVM opcodes and complete Rust programs, demonstrating the applicability of compiling by proving across different programming models.

\section{Related Work}
\label{sec:related-work}

\textbf{Verified Compilation.}
CompCert~\cite{Leroy2009CompCert} achieves verified compilation through 100,000 lines of Coq proofs over six person-years, manually proving each optimization correct. Recent systems like CakeML~\cite{Kumar2014CakeML}, Vélus~\cite{Bourke2017Velus}, and CertiCoq~\cite{Anand2017CertiCoq} extend verification to ML, Lustre, and Coq respectively, but still require manual optimization design. Translation validation~\cite{Pnueli1998,Necula2000TranslationValidation} and tools like ALIVE~\cite{lopesAlive2BoundedTranslation2021} verify optimizations post-hoc but cannot discover them. Our approach automatically extracts optimizations from semantic specifications—proof construction itself generates optimized rules, eliminating manual design.

\textbf{Proof-Producing Compilation.}
Classical PCC~\cite{Necula1997PCC} attaches safety proofs to compiled code. Modern systems like Vale~\cite{Bond2017Vale} and Jasmin~\cite{Almeida2020Jasmin} generate proofs alongside assembly code for cryptographic primitives, while EverCrypt~\cite{Protzenko2019EverCrypt} multiplexes verified implementations. These approaches still manually design optimizations then prove them correct. This work inverts this approach: proof construction automatically discovers and applies optimizations, making the proof itself the compilation mechanism.

\textbf{Program Optimization Techniques.}
Partial evaluation~\cite{JonesGomardSestoft1993} specializes programs through value propagation but requires manual binding-time analysis. Supercompilation~\cite{sorensen1994positive} performs aggressive optimization but faces exponential blowup. Superoptimization techniques like STOKE~\cite{Schkufza2013StochasticSuperoptimization} use MCMC to search for optimal x86 sequences, while SOUPER~\cite{Lopes2017Souper} synthesizes LLVM peephole optimizations through constraint solving. These approaches operate on concrete programs with manual guidance or search implementation spaces hoping to find optimizations, providing no coverage guarantees. The approach presented here operates on semantic specifications directly, using instruction boundaries to ensure tractability while achieving systematic coverage through exhaustive symbolic execution.

\textbf{Semantic Frameworks.}
Executable frameworks like $\K$~\cite{SerbanutaRosu2010K,ChenLucanuRosu2021b}, PLT Redex~\cite{Klein2010PLTRedex}, and Ott~\cite{Sewell2007Ott} generate tools from specifications but suffer performance penalties from modular rule decomposition. Verification frameworks like Dafny~\cite{Leino2010Dafny}, F*~\cite{Swamy2016FunctionalPrograms}, and Lean 4~\cite{deMoura2021Lean4} leave optimization as manual work. This approach bridges the gap: compiling modular specifications into monolithic optimized rules, achieving hand-optimized performance with formal guarantees.

\textbf{Blockchain Optimization.}
Gasol~\cite{Albert2023CAV} superoptimizes EVM bytecode for gas reduction. EthBMC~\cite{Frank2020EthBMC} performs bounded model checking. Solidity optimizes at source level. These target specific programs, not the VM itself. Our technique compiles EVM semantics directly, accelerating all tools built on those semantics—achieving ecosystem-wide speedups rather than per-program improvements.

\textbf{Program Synthesis.}
Synthesis approaches~\cite{Gulwani2017ProgramSynthesis,Solar-Lezama2008SketchSynthesis} generate programs through search or constraint solving, limited to small programs. Neural synthesis~\cite{Chen2021TransformersNeural} lacks formal guarantees. The presented approach synthesizes through deterministic proof compilation: systematically extracting all optimizations embedded in semantic specifications with machine-checkable correctness proofs.

Our \emph{compiling by proving} paradigm uniquely combines automatic optimization discovery, systematic coverage, and formal correctness—extracting optimizations from semantic specifications rather than designing them manually or searching for them heuristically. This delivers practical speedups while maintaining the correctness guarantees essential for compiler applications.

\section{Conclusion}
\label{sec:conclusion}

This paper presented \emph{compiling by proving}, a paradigm where constructing correctness proofs simultaneously generates optimized code. We showed that All-Path Reachability Proofs—built through symbolic execution to verify program properties—encode sufficient information for compilation. By applying graph transformations to these proof structures, we consolidate many modular semantic rules into single efficient rules, eliminating rewriting overhead while preserving correctness by construction.

Our language-agnostic implementation as a $\K$ framework extension demonstrates practical viability for different semantics. For EVM opcodes, we successfully compiled 131 of 149 opcodes, achieving consistent performance improvements across extensive conformance tests. For a Rust program via MIR semantics, whole-program compilation yielded considerably greater speedup, illustrating how benefits scale with compilation scope.

This work takes a concrete step toward unifying verification and optimization, as outlined in our introduction. Rather than treating proofs as mere correctness certificates, we showed they can drive compilation—the same symbolic execution that verifies program properties generates the optimized rules. This transforms the economics of formal methods: proof construction becomes a productive activity yielding both correctness guarantees and performance improvements, making formal verification practical for systems where both properties are essential.



\bibliographystyle{IEEEtran}
\bibliography{reference}
\appendix

\section{Detailed Opcode Compilation Results}
\label{sec:opcode-details}

Table~\ref{tab:step3_opcode_list} provides the complete breakdown of opcodes for which we successfully constructed program specifications versus those that remain future work.

\begin{table}[htbp]
\centering
\caption{Opcode Summarization Results}
\label{tab:step3_opcode_list}
\begin{tabularx}{\columnwidth}{lXX}
\toprule
Category & Successful Opcodes & Failed Opcodes \\
\midrule
Arith. \& Bit. & SUB, DIV, ADD, MOD, SMOD, SDIV, ADDMOD, MULMOD, SIGNEXTEND, AND, EVMOR, XOR, NOT, SHL, BYTE, SHR & MUL, EXP, SAR \\
\midrule
Comparison & LT, GT, SLT, EQ, SGT, ISZERO &  \\
\midrule
Flow Control & STOP, PC, JUMPDEST, INVALID, RETURN, REVERT, UNDEFINED & JUMP, JUMPI \\
\midrule
Stack & POP, PUSHZERO, PUSH1-32, DUP1-16, SWAP1-16 &  \\
\midrule
Memory & MLOAD, MSTORE, MSTORE8, MSIZE, MCOPY &  \\
\midrule
Storage & SLOAD, SSTORE &  \\
\midrule
Trans. Storage & TLOAD, TSTORE &  \\
\midrule
Environment & COINBASE, BLOCKHASH, TIMESTAMP, NUMBER, PREVRANDAO, DIFFICULTY, GASLIMIT, CHAINID & BASEFEE \\
\midrule
Context & ADDRESS, ORIGIN, CALLER, CALLDATALOAD, CALLDATASIZE, CALLDATACOPY, CODESIZE, CODECOPY, GASPRICE, BALANCE, RETURNDATASIZE, RETURNDATACOPY, SELFBALANCE, GAS & CALLVALUE, EXTCODECOPY, EXTCODEHASH, EXTCODESIZE \\
\midrule
System & LOG0-4 & SHA3 \\
\midrule
Contract &  & CALL, CALLCODE, CREATE, CREATE2, DELEGATECALL, SELFDESTRUCT, STATICCALL \\
\bottomrule
\end{tabularx}
\end{table}

\end{document}